\documentclass[12pt,english]{article}
\usepackage[LGR,T1]{fontenc}
\usepackage[latin1]{inputenc}
\usepackage{geometry}
\usepackage{amsmath}
\usepackage{amssymb}
\usepackage{graphicx}
\usepackage{esint}

\makeatletter

\DeclareRobustCommand{\greektext}{%
  \fontencoding{LGR}\selectfont\def\encodingdefault{LGR}}
\DeclareRobustCommand{\textgreek}[1]{\leavevmode{\greektext #1}}
\ProvideTextCommand{\~}{LGR}[1]{\char126#1}

\@ifundefined{date}{}{\date{}}

\usepackage{units}\usepackage{amstext}\usepackage{dsfont}\usepackage{setspace}\usepackage{mathrsfs}\usepackage{longtable}\usepackage{times}\usepackage{ulem}\usepackage{babel}

\newcommand{\suppress}[1]{}

\newlength\wvtextpercent
\newbox\strikebox
\def\strike#1{\setbox\strikebox \hbox{<#1>}\hbox{\raise0.5ex\hbox to 0pt{\vrule height 0.4pt width \wd\strikebox\hss}\copy\strikebox}}

\makeatother

\usepackage{babel}
\begin{document}


\linespread{1.5}
\title{\textbf{\Huge{}On the nature of the Born rule}}
\author{\textbf{Roland Riek} \\
 \textbf{Laboratory of Physical Chemistry, ETH Zurich, Switzerland}}

\maketitle
Keywords:

\vspace{3cm}

\linespread{1}

Address for correspondence:\\
 Roland Riek\\
 Laboratory of Physical Chemistry\\
 ETH Zuerich\\
 Wolfgang-Pauli-Strasse 10\\
 HCI F 225\\
 CH-8093 Zurich\\
 Tel.: +41-44-632 61 39\\
 e-mail: roland.riek@phys.chem.ethz.ch

\textbf{\large{}\newpage{} }{\large\par}

\section{Abstract}

A physical experiment comprises along the time trajectory a start,
a time evolution (duration), and an end, which is the measurement.
In non relativistic quantum mechanics the start of the experiment
is defined by the wave function at time 0 taking into account the
starting conditions, the evolution is described by the wave function
following the Schrödinger equation and the measurement by the Born
rule. While the Schrödinger equation is deterministic, it is the Born
rule that makes quantum mechanics statistical with all its consequences.
The nature of the Born rule is thereby unknown albeit necessary since
it produces the correct ensemble averaged measures of the experiment.
Here, it is demonstrated that the origin of the Born rule is the projection
from the quantum frame (i.e. wave description) to the classical mechanics
frame (i.e. particle description) described by a Ehrenfest theorem-oriented
Fourier transformation. The statistical averaging over many measurements
is necessary in order to eliminate the unknown initial and end time
coordinate of the experiment in reference to the beginning of the
universe. 

\section{Significance Statement}

One of the many counter-intuitive phenomena of quantum mechanics is
the loss of a deterministic description into a statistical classical
mechanics read-out by the measurement. The measurement appears thereby
to be the bridge between the quantum mechanics and classical mechanics
world mathematically formulated by the Born rule. However, the nature
of the Born rule is unknown. Here, it is demonstrated that the Born
rule originates from the experimental lack of knowledge on the time
coordinates at the start of the experiment, while the measurement
itself is a Fourier transformation of its kind from the quantum frame
(i.e. wave description) to the classical mechanics frame (i.e. particle
description) following the Ehrenfest theorem. This finding demystifies
the measurement problem with the Born rule and quantum mechanics in
general because it makes quantum mechanics genuinely deterministic
as classical mechanics is.

\section{Introduction}

The measurement of a quantum mechanical system can be considered a
transformation from the quantum mechanical description to its classical
mechanics one {[}1-5{]}. While both descriptions within are both deterministic
and time reversible (in quantum mechanics with the Schrödinger equation
and in classical mechanics with the Newtonian laws) it is stated that
both are lost with the measurement. Within the standard quantum mechanical
frame work this odd phenomenon under some conditions also called the
collapse of the wave function is usually explained by the involvement
of the observer (i.e. measurement device) in the experiment, a requested
quantum mechanical super position between measurement device and system
under study, entanglement between the system under study and the environment,
by decoherence due to an interaction with the environment, or by our
apparent limitation to be able to detect only at the classical limit
(i.e. being unable to measure quantum mechanically) {[}6-12{]}. It
also nourishes distinct ontologies of quantum mechanics starting from
the Copenhagen interpretation, via the Bohm-de Broglie, and Everett's
many world to retrocausal interpretations and discrete approaches
{[}13-24{]}. It further builds the basis for the non-locality of quantum
mechanics highlighted prominently by the Einstein-Podolski-Rosen paradox
(EPR) in combination with the Bell inequalities {[}25-30{]}. 

The mathematical description of the measurement is the Born rule {[}1-5{]}.
It yields statistically the correct result of the experiment, but
can not calculate with certainty the result of a single experiment.
Einstein and others concluded therefore that the established formulation
of quantum mechanics must be incomplete {[}25{]}. However, the success
of quantum mechanics and the experimental evidence collected on the
Bell inequalities as well as the meausrement described by a decoherence
phenomenon are in favor of quantum mechanics as is {[}26-32{]}. Nonetheless,
due to the odd properties of quantum mechanics extensions thereof
or other approaches are still discussed {[}for example 13-24{]} and
in light of the fundamental inconsistencies between quantum mechanics
and general relativity requested {[}33{]}. 

In the work presented, the measurement problem of an experiment is
revisited. It is thereby assumed that the measurement itself is a
transformation from the wave to a particle description by a Fourier
transformation following the Ehrenfest theorem, which is deterministic.
The statistical origin of the Born rule is due to the time unknown
nature of the initial $t_{i}$ and end $t_{f}$ time point of the
experiment in relation to the absolute start of time, which varies
from measurement to measurement, while the experimental time is known
$t_{e}=t_{f}-t_{i}$. By doing so, the formulation presented yields
for each experiment a deterministic result which converges to the
Born rule upon statistically averaging. It is thereby demonstrated
that the statistical origin of the Born rule is the unknown initial
and end time coordinates of the experiment, while nature is deterministic
both at the quantum mechanical as well as at the classical mechanics
level including the measurement. 

After a short summary on useful standard quantum mechanics (4.1),
a single measurement is studied by assuming a frame change from the
quantum mechanics frame to the classical mechanics frame by a Fourier
transformation (4.2), followed in 4.3 by the description of the time
evolution of a quantum mechanical system from its start to the measurement.
In (5) the results are discussed.

\newpage{}

\section{Theory}

\subsection{Standard Quantum Mechanics }

Within the non relativistic quantum mechanics the time evolution $t$
of the wave function $\text{\textgreek{Y}}(\vec{r},\,t)$ describing
the system under study (with $\vec{r}$ being the space vector) is
described by the time-dependent Schrödinger equation 

\begin{equation}
\hat{H}\:\text{\textgreek{Y}}(\vec{r},\,t)=i\:\hbar\frac{\partial}{\partial t}\text{\textgreek{Y}}(\vec{r},\,t)
\end{equation}

with $\hat{H}=-\frac{\hbar^{2}}{2m}\nabla^{2}+V(\vec{r})$ the Hamilton
operator (with $m$ the mass of the particle and $V(\vec{r})$ the
acting potential), $\hbar$ the reduced Planck constant with $\hbar=h/2\pi$
, and $i=\sqrt{-1}$ . In the following description translation along
the time (by $\delta$) is required. The following transformation
is then given:

\begin{equation}
\text{\textgreek{Y}}(\vec{r},\,t-\delta)=\hat{U}_{t}(\delta)\:\text{\textgreek{Y}}(\vec{r}\,,\,t)
\end{equation}

with the unitary operator $\hat{U}_{t}(\delta)=e^{\frac{i}{h}\delta\:\hat{E}}$
(with the energy tensor $\hat{E}=i\hbar\frac{\text{\ensuremath{\partial}}}{\text{\ensuremath{\partial}}t}$
).

In the following the Hamilton operator is considered time-independent
yielding 

\begin{equation}
\text{\textgreek{Y}}(\vec{r},\,t)=\:e^{-\frac{i}{\hbar}\hat{H}\,t}\text{\textgreek{Y}}(\vec{r},\,0_{e})
\end{equation}
 enabling a separation between space and time with time starting at
$0$ (denoted $0_{e}$) for the beginning of the experiment. The solutions
of the Schrödinger equation are then given by 

\begin{equation}
\text{\textgreek{Y}}_{n}(\vec{r},\,t)=\text{\textgreek{Y}}_{n}(\vec{r\,})\:e^{-\frac{i}{\hbar}E_{n}\,t}
\end{equation}
with $n$ an integer, and $\hat{H}\,\text{\textgreek{Y}}_{n}(\vec{r})=E_{n}\text{\textgreek{Y}}_{n}(\vec{r})$
with $E_{n}=const.$ (i.e. $E_{n}=\hbar\,\omega_{n}$). 

The general solution of the Schrödinger equation is then given by
the superposition of all the $\text{\textgreek{Y}}_{n}(\vec{r},\,t)$:

\begin{equation}
\text{\textgreek{Y}}(\vec{r},\,t)=\sum_{n}c_{n}(0_{e})\text{\textgreek{Y}}_{n}(\vec{r},\,t)=\sum_{n}c_{n}(0_{e})\text{\textgreek{Y}}_{n}(\vec{r\,})\:e^{-\frac{i}{\hbar}E_{n}\,t}
\end{equation}
with $c_{n}(0_{e})=\int\text{\textgreek{Y}}(\vec{r},\,0_{e})\,\text{\textgreek{Y}}_{n}^{*}(\vec{r},\,0_{e})\,dV\,\geq0$
and $\text{\textgreek{Y}}_{n}(\vec{r},\,t)$, which are orthonormal
to each other (i.e. $\text{<\textgreek{Y}}_{n}|\,\text{\textgreek{Y}}_{m}>=\delta_{mn}$
using the Dirac notation and with $\delta_{mn}$ being the Kronecker's
symbol). 

$c_{n}(0_{e})$ is thus independent of time and correspondingly $c_{n}(0_{e})=c_{n}(0)$.

The mean value of a measurement on an observable described by the
hermitian operator $\hat{A}$ is described by the Born rule 

\begin{equation}
<\hat{A}>=<\text{\textgreek{Y}(\ensuremath{\vec{r}},\,t)}|\hat{A|}\text{\textgreek{Y}(\ensuremath{\vec{r}},\,t)}>=\int\text{}\text{\textgreek{Y}}(\vec{r},\,t)^{*}\hat{A}\,\text{\textgreek{Y}}(\vec{r},\,t)\,dV=\int\text{}\varPhi(\vec{r},\,t)^{*}\hat{A}\,\text{\ensuremath{\varPhi}}(\vec{r},\,t)\,dV
\end{equation}

with $\varPhi(\vec{r},\,t)=\sum_{n}b_{n}(0)\text{\ensuremath{\varPhi}}_{n}(\vec{r},\,t)$
if the set of $\text{\ensuremath{\varPhi}}_{n}(\vec{r})$ is a complete
orthonormal system of the operator $\hat{A}$ with the classical observables
$a_{m}$ (which are the eigenvalues of the operator) and with $b_{n}(0)=\int\text{\ensuremath{\varPhi}}(\vec{r},\,0)\,\text{\ensuremath{\varPhi}}_{n}^{*}(\vec{r},\,0)\,dV\,$

\begin{equation}
<\hat{A}>=\sum_{m}|b_{m}^{*}b_{m}|\:a_{m}
\end{equation}
 with $|b_{m}^{*}b_{m}|=|b_{m}^{2}|$ the probability that the value
$a_{m}$ is measured. In particular, $|\text{\ensuremath{\varPhi}}_{n}(\vec{r\,})^{*\,}\text{\ensuremath{\varPhi}}_{n}(\vec{r\,})|$
describes the probability at position $\vec{r\,}$. If $\hat{A}$
commutes with $\hat{H}$, $\text{\ensuremath{\varPhi}}_{n}(\vec{r},\,t)=\,\text{\textgreek{Y}}_{n}(\vec{r},\,t)$
can be selected and $c_{n}(0_{e})=b_{n}(0_{e})$, respectively.

\subsection{Revisiting the concept of a measurement and an experiment}

In contrast to standard quantum mechanics, we define the measurement
as a a change in the reference frame from a wave description following
quantum mechanics to a particle/state description following classical
mechanics orchestrated by a Fourier transformation of its kind following
the Ehrenfest theorem. Furthermore the read out should comprise the
entire measured information of the system under investigation which
is preserved by the Fourier transformation. Let us describe this Ansatz
by a classical analog of a sound wave $\varphi(t)=cos(\omega\,t)$
with its Fourier transform analog to be $\omega$(if infinitely long
investigated) capturing the entire information. While there is a sound
wave we hear the frequency $\omega$. It is important to note that
theoretically the Fourier transformation guarantees that no information
is lost in the measurement and thus both descriptions are equivalent
(please note, that a loss of information happens however due to the
timing issues discussed below). With other words, the event of a single
measurement at time $t_{f}$ ($f$ for final, and with the big bang
as the reference time $t=0$) is regarded and defined here as a projection
from the wave description of a quantum mechanical system to the particle
presentation through a Fourier transformation without loss of information.
For the description of such a measurement a system $\text{\textgreek{Y}}(\vec{r},\,t)$
is selected that evolves under a non time-dependent Hamilton and that
the observable of interest is described by the hermitian operator
$\hat{A}$ yielding $\digamma\hat{A}\text{\textgreek{Y}}$ with $\digamma$
being the Fourier transformation (please note, the normalisation of
the Fourier transformation is omitted here).

In the most simple case of a free particle this reads as 

\begin{equation}
\digamma\hat{x}\text{\textgreek{Y}}(\vec{r},\,t_{f})=\int dVe^{-\frac{i}{\hbar}\vec{r}\:\vec{p}}\hat{x}\text{\textgreek{Y}}(\vec{r},\,t_{f})
\end{equation}

if the position of the particle is of interest, while 

\begin{equation}
\digamma\hat{p}\text{\textgreek{Y}}(\vec{p},\,t_{f})=\int dp^{3}e^{-\frac{i}{\hbar}\vec{r}\:\vec{p}}\hat{p}\text{\textgreek{Y}}(\vec{p},\,t_{f})
\end{equation}

if the momentum is of interest. Please note, that both expressions
are for a single measurement and not yet comparable with the equivalent
Born descriptions $<\hat{x}>$ and $<\hat{p}>$, which is a statistical
average. 

It is critical to note here, that when the experiment is repeated
$t_{f}$ changes because time is ongoing with the origin of time assumed
to be at the bing bang. Obviously this contrasts standard quantum
mechanics, which start of each experiment is defined as $t_{i}=0$
and the duration of the experiment is given by a defined value $t_{e}$
with $t_{f}=t_{e}$ and thus $t_{f}$ is the same with each experiment.
Under the assumption of an ongoing time with each experiment having
a different $t_{i}$ and $t_{f}$ but the same experimental time $t_{e}=t_{f}-t_{i}$
the evolution of the periodic wave function can be rewritten with
the orthonormal basis $\text{\textgreek{Y}}_{n}$ into 
\begin{equation}
\text{\textgreek{Y}}(\vec{r}\,,t_{f})=\text{\textgreek{Y}}(\vec{r},\,t_{i}+t_{e})=\text{\ensuremath{\hat{U}_{t}}}(-t_{i})\text{\textgreek{Y}}(\vec{r},\,t_{e})=\sum_{n}c_{n}(t_{i})\text{\ensuremath{\hat{U}_{t}^{n}}}(-t_{i})\text{\textgreek{Y}}_{n}(\vec{r},\,t_{e})
\end{equation}
with $\hat{U}_{t}^{n}(-t_{i})=e^{-\frac{i}{h}t_{i}\:E_{n}}$ yielding
with a time-independent Hamiltonian

\begin{equation}
\text{\textgreek{Y}}(\vec{r}\,,t_{f})=\sum_{n}c_{n}(t_{i})\,e^{-\frac{i}{\hbar}t_{i}\:E_{n}}\Psi_{n}(\vec{r},\,t_{e})
\end{equation}

In a first step towards the measurement the observable operator is
added at time point $t_{f}$. 

\begin{equation}
\hat{A}\text{\textgreek{Y}}(\vec{r}_{f}\,,t_{f})=\sum_{n}c_{n}(t_{i})\,\hat{A}\,e^{-\frac{i}{\hbar}t_{i}\:E_{n}}\Psi_{n}(\vec{r},\,t_{e})=\sum_{n}c_{n}(t_{i})\,e^{-\frac{i}{\hbar}t_{i}\:E_{n}}\hat{A}\text{\textgreek{Y}}_{n}(\vec{r},\,t_{e})
\end{equation}
 Eq. 10 indicates that due to the start of the experiment with $t_{i}$
each orthonormal state described with $\text{\textgreek{Y}}_{n}$
has its own phase (i.e. $-\frac{i}{h}t_{i}\:E_{n}$ in the exponent)
and by a repetition of the experiment this phase alters because the
initial time $t_{i}$ alters every time. Thus, albeit deterministically
a repetition of the experiment does not yield the same wave function
after evolution during time $t_{e}$. This interpretation requests
the definition of a statistical measure of the observable to be dependent
on available information-only, which is $t_{e}$ as given successfully
(while ad hoc) by the Born rule (i.e. $<\hat{A}>=\int\text{\textgreek{Y}}(\vec{r},\,t_{e})^{*}\hat{A}\,\text{\textgreek{Y}}(\vec{r},\,t_{e})\,dV=\int\text{}\varPhi(\vec{r},\,t)^{*}\hat{A}\,\text{\ensuremath{\varPhi}}(\vec{r},\,t)\,dV=\sum_{n}|b_{n}^{*}b_{n}|\:a_{n}$).

Hence, in the next steps the dependence on the start of the experiment
with $t_{i}$ is described by $t_{e}$ . This is possible since the
exponents in eq. 10 are of periodic nature with time periodicity $\tau_{n}=\frac{h}{E_{n}}=\frac{2\pi\,\hbar}{E_{n}}$
such that

\begin{equation}
t_{i}=\alpha_{n,i}\tau_{n}+\triangle t_{n,i}
\end{equation}

and 
\begin{equation}
t_{e}=\alpha_{n,e}\tau_{n}+\triangle t'_{n,e}+\triangle t_{n,e}^{''}
\end{equation}

with $\alpha$'s integers and $0\leq\triangle t_{n,i}\leq\tau_{n}$
, $0\leq\triangle t_{n,e}^{'}+\triangle t_{n,e}^{''}\leq\tau_{n}$
(Figure 1). It is

\begin{equation}
\triangle t_{n,i}+\triangle t_{n,e}^{'}=\tau_{n}
\end{equation}

Next, these relations are expanded to $t_{i}$ and $t_{e}$ using
the periodicity of the wave function

\begin{equation}
\triangle t_{n,i}=\tau_{n}-\triangle t_{n,e}^{'}
\end{equation}

\begin{equation}
t_{i}=\triangle t_{n,i}+\alpha_{n,i}\tau_{n}=\tau_{n}-\triangle t_{n,e}^{'}-(\alpha_{n,e}+1)\tau_{n}=-t_{e}+\triangle t_{n,e}^{''}
\end{equation}

without loss of information within the formulas needed to describe
an experiment. This leads to

\begin{equation}
\hat{A}\text{\textgreek{Y}}(\vec{r}\,,t_{f})=\sum_{n}c_{n}(t_{i})\,e^{-\frac{i}{\hbar}\triangle t_{n,e}^{''}\:E_{n}}\,\,e^{+\frac{i}{\hbar}t_{e}\:E_{n}}\hat{A}\text{\textgreek{Y}}_{n}(\vec{r},\,t_{e})
\end{equation}

with the following further boundary conditions (with $\triangle t_{n,e}^{''}<\tau_{n}-\triangle t_{n,e}^{'}$
and $\triangle t_{n,i}+\triangle t_{n,e}^{'}=\tau_{n}$):
\begin{equation}
\;\triangle t_{n,e}^{''}<\triangle t_{n,i}<\tau_{n}\;
\end{equation}

With other words, from experiment to experiment $\hat{A}\text{\textgreek{Y}}(\vec{r}\,,t_{f})$
varies because $\triangle t_{n,e}^{''}$ varies.

Next, we need to resolve the potential $t_{i}$ dependency of the
coefficients $c_{n}(t_{i})$. In the standard description of quantum
mechanics $c_{n}(t=0)$ and thus independent of time, while the experimental
time dependency of the system is only within the wave function. This
is also true for the presented discussion since $c_{n}(t_{i})=\int\text{\textgreek{Y}}(\vec{r},\,t_{i})\,\text{\textgreek{Y}}_{n}^{*}(\vec{r},\,t_{i})\,dV=\int\text{\textgreek{Y}}(\vec{r},\,0)\,\text{\textgreek{Y}}_{n}^{*}(\vec{r},\,0)\,dV$,
which yields $c_{n}(t_{i})\equiv c_{n}'$ and thus is time independent.

Eq. 18 reads now 

\begin{equation}
\hat{A}\text{\textgreek{Y}}(\vec{r}\,,t_{f})=\sum_{n}c'_{n}\,e^{-\frac{i}{\hbar}\triangle t_{n,e}^{''}\:E_{n}}\,\,e^{+\frac{i}{\hbar}t_{e}\:E_{n}}\hat{A}\text{\textgreek{Y}}_{n}(\vec{r},\,t_{e})
\end{equation}

When averaged over many experiments the following is obtained by integrating
over $\triangle t_{n,i}$ from $\frac{-\tau_{n}}{2}$ to $\frac{\tau_{n}}{2}$.
These integration borders are necessary in order to have both the
mean $\triangle t_{n,i}$ and the mean $t_{i}$ modulo $\tau_{n}$
equals to $0$ (please note, an integration from $-\tau_{n}$ to $\tau_{n}$
would result in mean $\triangle t_{n,i}$ and the mean $t_{i}$ modulo
$\tau_{n}$ of $\frac{\tau_{n}}{2}$ and would result in a non exactly
defined $t_{i}$ as each $\tau_{n}$ is different).

\begin{equation}
\{\hat{A}\text{\textgreek{Y}}(\vec{r}\,,t_{f})\}_{time\,average}=\sum_{n}c'_{n}\,\frac{1}{\tau_{n}}\intop_{-\tau_{n}/2}^{\tau_{n}/2}d\triangle t_{n,i}\frac{1}{\triangle t_{n,i}}\intop_{0}^{\triangle t_{n,i}}d\triangle t_{n,e}^{''}e^{-\frac{i}{\hbar}\triangle t_{n,e}^{''}\:E_{n}}\,\,e^{+\frac{i}{\hbar}t_{e}\:E_{n}}\hat{A}\text{\textgreek{Y}}_{n}(\vec{r},\,t_{e})
\end{equation}

\begin{equation}
\{\hat{A}\text{\textgreek{Y}}(\vec{r}\,,t_{f})\}_{time\,average}=\sum_{n}c'_{n}\frac{1}{\pi}\,\intop_{0}^{\pi}\,\frac{sin(t)}{t}dt\,e^{+\frac{i}{\hbar}t_{e}\:E_{n}}\hat{A}\text{\textgreek{Y}}_{n}(\vec{r},\,t_{e})
\end{equation}

with $\frac{1}{\pi}\intop_{0}^{\pi}\,\frac{sin(t)}{t}dt$ having a
real value and can be absorbed into the coefficients $c'_{n}\frac{1}{\pi}\intop_{0}^{\pi}\,\frac{sin(t)}{t}dt\,=c_{n}(0)$.
Please also note, that the normalisation factor for the Fourier transformation
could be incoporated here as well.

\begin{equation}
\{\hat{A}\text{\textgreek{Y}}(\vec{r}\,,t_{f})\}_{time\,average}=\sum_{n}c{}_{n}(0)\,\,e^{+\frac{i}{\hbar}t_{e}\:E_{n}}\hat{A}\text{\textgreek{Y}}_{n}(\vec{r},\,t_{e})
\end{equation}

The Fourier transformation of the measurement is applied to the system.
In the most simple case of a free particle's position this reads as 

\begin{equation}
\digamma\{\hat{x}\text{\textgreek{Y}}(\vec{r},\,t_{f})\}_{time\,average}=\int dVe^{-\frac{i}{\hbar}\vec{r}\:\vec{p}}\sum_{n}c{}_{n}(0)\,\,e^{+\frac{i}{\hbar}t_{e}\:E_{n}}\hat{x}\text{\textgreek{Y}}_{n}(\vec{r},\,t_{e})
\end{equation}
\[
=\int dV\text{\textgreek{Y}}^{*}(\vec{r},\,0)\,\sum_{n}\,e^{+\frac{i}{\hbar}t_{e}\:E_{n}}\hat{x}c{}_{n}(0)\text{\textgreek{Y}}_{n}(\vec{r},\,t_{e})=\int dV\sum_{m}\sum_{n}c{}_{m}^{*}(0)\text{\textgreek{Y}}_{m}^{*}(\vec{r},\,0)\,\,e^{+\frac{i}{\hbar}t_{e}\:E_{n}}\hat{x}c{}_{n}(0)\text{\textgreek{Y}}_{n}(\vec{r},\,t_{e})
\]

\[
=\int dV{\displaystyle \sum_{n}}c{}_{n}^{*}(0)\text{\textgreek{Y}}_{n}^{*}(\vec{r},\,0)\,\,e^{+\frac{i}{\hbar}t_{e}\:E_{n}}\hat{x}c{}_{n}(0)\text{\textgreek{Y}}_{n}(\vec{r},\,t_{e})=<\text{\textgreek{Y}}(\vec{r},\,t_{e})|\hat{x}|\text{\textgreek{Y}}(\vec{r},\,t_{e})>=<\hat{x}>
\]

because $\text{<\textgreek{Y}}_{n}|\,\text{\textgreek{Y}}_{m}>=\delta_{mn}$,
$\hat{x}$ and $\text{\textgreek{Y}}_{n}$ commute and $\text{\textgreek{Y}}^{*}(\vec{r},\,0)=e^{-\frac{i}{\hbar}\vec{r}\:\vec{p}}$.
Thus, the combination of the Fourier transformation with experimental
repetition yielding time averaging yields the Born rule for the simple
case discussed. 

For a more general case with $\text{\textgreek{F}}(\vec{r},\,t_{f})=\sum_{m}b_{m}(0)\text{\textgreek{F}}_{m}(\vec{r},\,t_{f})$
with $\text{\textgreek{F}}_{m}$ being the orthonormal basis of $\hat{A}$
(with $b_{m}(0)\:=\intop\text{\textgreek{F}}(\vec{r},\,0)\text{\textgreek{F}}_{m}^{*}(\vec{r},\,0)\,dV$)

the Fourier transformation for the measurement is then given by

\begin{equation}
\digamma\{\hat{A}\text{\textgreek{F}}(\vec{r},\,t_{f})\}=\int dV\,\text{\textgreek{F}}^{*}(\vec{r},\,0)\,\hat{A}\text{\textgreek{F}}(\vec{r},\,t_{f})=\int dV\,\sum_{m}b_{m}^{*}(0)\text{\textgreek{F}}_{m}^{*}(\vec{r},\,0)\,\hat{A}\,\sum_{k}b_{k}(0)\text{\textgreek{F}}_{k}(\vec{r},\,t_{f})
\end{equation}

with $\hat{A}\,\text{\textgreek{F}}_{k}=a_{k}\text{\textgreek{F}}_{k}$
due to the orthonormal property of $\text{\textgreek{F}}_{k}$ in
respect to $\hat{A}\,$.

\begin{equation}
\digamma\{\hat{A}\text{\textgreek{F}}(\vec{r},\,t_{f})\}=\int dV\,\sum_{m}b_{m}^{*}(0)\text{\textgreek{F}}_{m}^{*}(\vec{r},\,0)\,\,\sum_{k}a_{k}\,b_{k}(0)\text{\textgreek{F}}_{k}(\vec{r},\,t_{f})
\end{equation}

with $\text{\ensuremath{\varPhi}}_{k}(\vec{r},\,t)=\sum_{l}\text{<\textgreek{Y}}_{l}(\vec{r},\,t)|\text{\ensuremath{\varPhi}}_{k}(\vec{r},\,t)>\text{\textgreek{Y}}_{l}(\vec{r},\,t)$

\begin{equation}
\digamma\{\hat{A}\text{\textgreek{F}}(\vec{r},\,t_{f})\}=\int dV\,\sum_{m}b_{m}^{*}(0)\sum_{l}\text{<\textgreek{Y}}_{l}^{*}(\vec{r},\,0)|\text{\ensuremath{\varPhi}}_{m}^{*}(\vec{r},\,0)>\text{\textgreek{Y}}_{l}^{*}(\vec{r},\,0)\,\,
\end{equation}
\[
\sum_{k}a_{k}\,b_{k}(0)\sum_{s}\text{<\textgreek{Y}}_{s}(\vec{r},\,t_{f})|\text{\ensuremath{\varPhi}}_{k}(\vec{r},\,t_{f})>\text{\textgreek{Y}}_{s}(\vec{r},\,t_{f})
\]

\begin{equation}
\digamma\{\hat{A}\text{\textgreek{F}}(\vec{r},\,t_{f})\}=\int dV\,\sum_{m}b_{m}^{*}(0)\sum_{l}\text{<\textgreek{Y}}_{l}^{*}(\vec{r},\,0)|\text{\ensuremath{\varPhi}}_{m}^{*}(\vec{r},\,0)>\text{\textgreek{Y}}_{l}^{*}(\vec{r},\,0)\,\,
\end{equation}
\[
\sum_{k}a_{k}b_{k}(0)\sum_{s}\text{\ensuremath{e^{+\frac{i}{\hbar}t_{e}\:E_{s}}e^{-\frac{i}{\hbar}\triangle t_{s,e}^{''}\:E_{s}}}<\textgreek{Y}}_{s}(\vec{r},\,t_{f})|\text{\ensuremath{\varPhi}}_{k}(\vec{r},\,t_{f})>\text{\textgreek{Y}}_{s}(\vec{r},\,t_{e})
\]

because of $\text{<\textgreek{Y}}_{n}|\,\text{\textgreek{Y}}_{m}>=\delta_{mn}$
$l=s$, which yields

\begin{equation}
\digamma\{\hat{A}\text{\textgreek{F}}(\vec{r},\,t_{f})\}=\int dV\,\sum_{m}\sum_{k}b_{m}^{*}(0)a_{k}b_{k}(0)\sum_{l}\text{<\textgreek{Y}}_{l}^{*}(\vec{r},\,0)|\text{\ensuremath{\varPhi}}_{m}^{*}(\vec{r},\,0)>\,
\end{equation}
\[
\text{<\ensuremath{\text{\ensuremath{\varPhi}}_{k}^{*}}(\ensuremath{\vec{r}},\,\ensuremath{t_{f}})|\textgreek{Y}}_{l}^{*}(\vec{r},\,t_{f})>\text{\textgreek{Y}}_{l}^{*}(\vec{r},\,t_{e})\text{\textgreek{Y}}_{l}(\vec{r},\,t_{e})\,e^{-\frac{i}{\hbar}\triangle t_{le}^{''}\:E_{l}}
\]

Because of the unitary property of the time operator with $<\varPhi(0)|\text{\textgreek{Y}}(0)>=<\varPhi(0)|\hat{U}_{t}^{+}(\delta)\hat{U}_{t}(\delta)|\text{\textgreek{Y}}(0)>=<\varPhi(\delta)|\text{\textgreek{Y}}(\delta)>$
eq. 30 can be simplified to 

\begin{equation}
\digamma\{\hat{A}\text{\textgreek{F}}(\vec{r},\,t_{f})\}=\int dV\,\sum_{m}\sum_{k}b_{m}^{*}(0)a_{k}b_{k}(0)\sum_{l}\text{<\textgreek{Y}}_{l}^{*}(\vec{r},\,0)|\text{\ensuremath{\varPhi}}_{m}^{*}(\vec{r},\,0)>\text{<\ensuremath{\text{\ensuremath{\varPhi}}_{k}^{*}}(\ensuremath{\vec{r}},\,0)|\textgreek{Y}}_{l}^{*}(\vec{r},\,0)>
\end{equation}
\[
\text{\textgreek{Y}}_{l}^{*}(\vec{r},\,t_{e})\text{\textgreek{Y}}_{l}(\vec{r},\,t_{e})\,e^{-\frac{i}{\hbar}\triangle t_{le}^{''}\:E_{l}}\,
\]

which for each measurement differs and can not be simplified further
as $\triangle t_{le}^{''}$ is different for each $l$ and changes
from measurement to measurement. 

After time averaging (see above and by incorporating the time averaging
value $\frac{1}{\pi}\intop_{0}^{\pi}\,\frac{sin(t)}{t}dt\,$ into
the term $b_{m}^{*}(0)b_{m}(0)$) we obtain 

\begin{equation}
\digamma\{\hat{A}\text{\textgreek{F}}(\vec{r},\,t_{f})\}_{time\,average}=\int dV\,\sum_{m}\sum_{k}b_{m}^{*}(0)a_{k}b_{k}(0)\sum_{l}\text{<\textgreek{Y}}_{l}^{*}(\vec{r},\,0)|\text{\ensuremath{\varPhi}}_{m}^{*}(\vec{r},\,0)>
\end{equation}
\[
\text{<\ensuremath{\text{\ensuremath{\varPhi}}_{k}^{*}}(\ensuremath{\vec{r}},\,0)|\textgreek{Y}}_{l}^{*}(\vec{r},\,0)>\text{\textgreek{Y}}_{l}(\vec{r},\,t_{e})\text{\textgreek{Y}}_{l}^{*}(\vec{r},\,t_{e})\,\,
\]

\begin{equation}
\digamma\{\hat{A}\text{\textgreek{F}}(\vec{r},\,t_{f})\}_{time\,average}=\,\sum_{m}b_{m}^{*}(0)a_{m}b_{m}(0)\,=<\hat{A}>
\end{equation}

\includegraphics[scale=0.4]{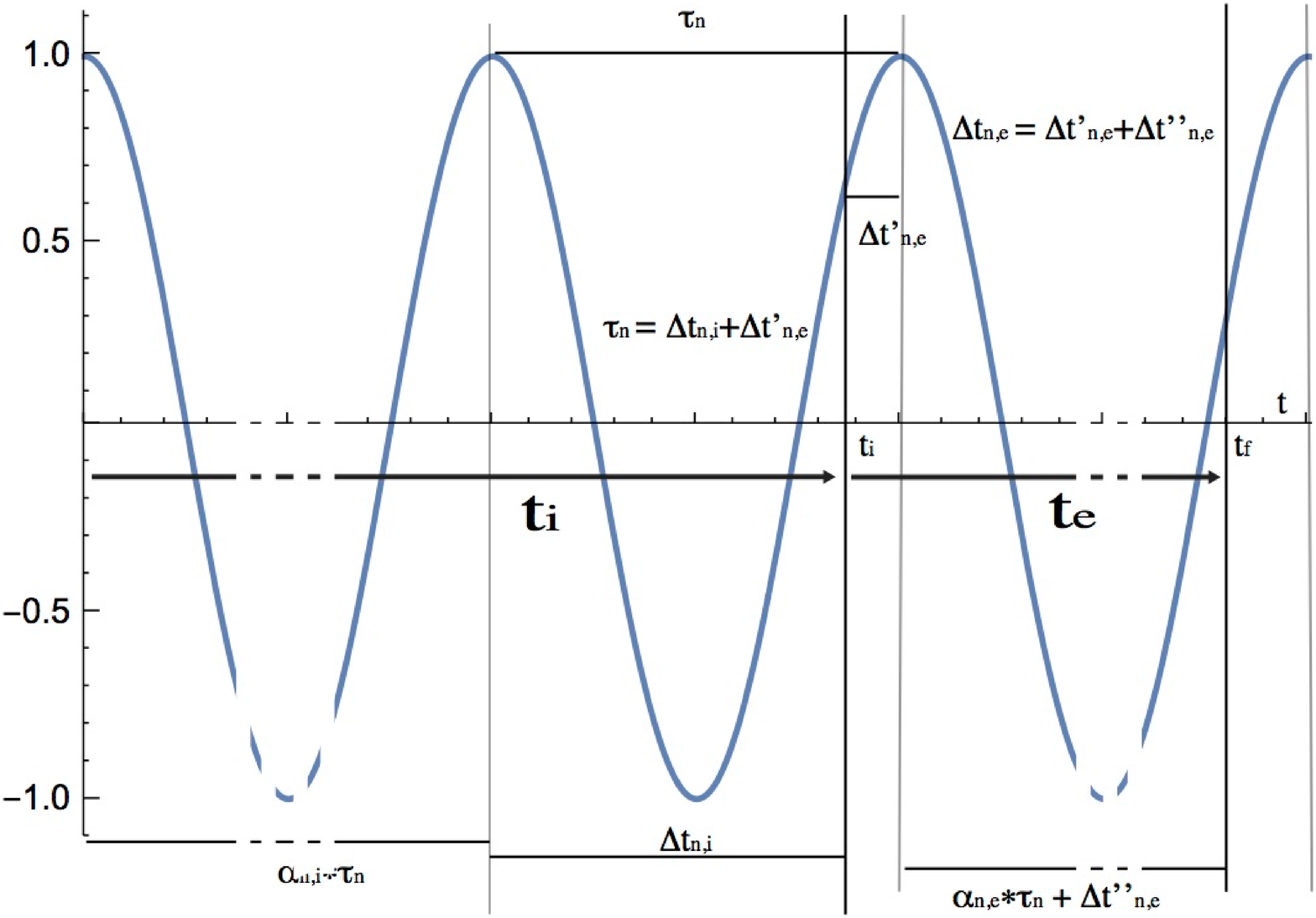}

Figure 1: The dependence between the various $\triangle$'s are illustrated
here for the time with $\triangle t_{n,e}=\triangle t_{n,e}^{'}+\triangle t_{n,e}^{''}<\tau$
and $\tau_{n}=\triangle t_{n,e}^{'}+\triangle t_{n,i}$ within a periodic
wave function (i.e. $cos(t/\tau_{n})$). 

\section{Discussion}

Assuming that the measurement of a (quantum mechanical) system is
a Ehrenfest theorem-type Fourier transformation and that time is universal
starting with time 0 at the beginning of the universe, the Born rule
has been derived here. It thereby not only links the wave and particle
description in quantum mechanics with its classical analog, but highlights
the origin of the statistical nature of the Born rule and thus quantum
mechanics to be the unknown absolute starting time $t_{i}$ and ending
time $t_{f}$ of an experiment, that change with each experiment.
Thus, quantum mechanics including the measurement is genuinly deterministic,
but not physical experiments as they have to be repeated and must
be repeatable while the experimentor lacks information on the absolute
time. While experimental physics is therefore restrained to the Born
rule, nature is not and thus deterministic both at the quantum mechanical
as well as at the Newtonian level and connected by a Fourier transformation.
Interestingly, if experiments can be designed with $t_{e}<<\tau$
then the quantum mechanical system should behave deterministic. With
other words, with increasing the time resolution of experiments, quantum
mechanics may get deterministic again to be demonstrated. In return,
a classical object can be regarded having a $\tau\thickapprox0$ (or
$\triangle t_{n,e}^{''}\approx const$ and $\triangle t_{n,i}\approx const$
for all $n$) and thus no statistical averaging is needed, which yields
also a deterministic result. While these claims may be valuable for
experimental physics, the presented approach to derive the origin
of the Born rule is also relevant for the ontology of physics opening
another approach than the Copenhagen interpretation, the Bohm-de Broglie,
Everett's many world as well as retrocausal interpretations of quantum
mechanics {[}13-24{]} by giving the rather obscure entity and variable
time more weight.

\section{Acknowledgment}

We would like to thank the ETH for unrestrained financial support
and Dr. Gianni Blatter, Dr. Alexander Sobol, Dr. Matthias Ernst, and
Dr. Witek Kwiatkowski for helpful discussions.

\section{References}

1. Greiner W, Neise L, Stöcke H 2001 \textit{Quantenmechanik} Verlag
Harri Deutsch: Frankfurt am Main. ISBN 3-540-67458-6

2. Landau LD, Lifshitz EM 1965 \textit{Quantum Mechanics} (Volume
3 of A Course of Theoretical Physics), Pergamon Press 1965

3. Schrödinger E, Born M 1935 Discussion of probability relations
between separated systems. \textit{Mathematical Proceedings of the
Cambridge Philosophical Society} 31, 555\textendash 563.

4. Born M 1926 Quantenmechanik der Stoßvorgänge. \textit{Z. Phys.}
38, 803-827.

5. Bohr N 1935 Can quantum-mechanical description of physical reality
be considered complete? \textit{Phys. Rev. }48:696\textendash 702,
doi:10.1103/PhysRev.48.696.

6. Ghirardi GC, Rimini A, Weber T 1986 Unified dynamics for microscopic
and macroscopic systems. \textit{Phys. Rev. D} 34:470-491.

7. Penrose R 2000 Wavefunction collapse as a real gravitational effect.
In: Fokas, A., Grigoryan, A., Kibble, T., Zegarlinski, B. (eds.) \textit{Mathematical
Physics 2000}, pp. 266-282. Imperial College, London.

8. Hepp K 1972 Quantum theory of measurement and macroscopic observables.
\textit{Helv. Phys. Acta} 45:237-248.

9. Fröhlich J, Schubnel B 2016 The preparation of states in quantum
mechanics. \textit{J. Math. Phys.} 57:042101

10. Fröhlich J, Schubnel B 2015 Quantum probability theory and the
foundations of quantum mechanics. \textit{Lecture Notes in Physics},
899:131-193.

11. Blanchard P, Fröhlich J, Schubnel B 2016 A \textquotedbl garden
of forking paths\textquotedbl{} - The quantum mechanics of histories
of events. \textit{Nuclear Physics} B 912:463-484.

12. Zurek WH 2003 Decoherence, einselection, and the quantum origins
of the classical. ArXiv:quant-ph/015127v3

13. Bohm D 1952 A Suggested Interpretation of the Quantum Theory in
Terms of 'Hidden Variables' I. \textit{Physical view} 85: 66\textendash 179.

14. Bopp FW 2018 Causal Classical Phyiscs in Time Symmetric Quantum
Mechanics. \textit{Proceedings }2:157doi: 10.3390/ecea-4-05-1.

15. Price H 2012 Does time-symmetry imply retrocausality? How the
quantum world says \textquotedbl Maybe\textquotedbl ? \textit{St.
Hist. Philos. Mod. Phys}. 43:75-83. 

16. Price H 1994 A neglected route to realism about quantum mechanics.
\textit{Mind} 103:303-336. 

17. H. Everett 1957 \textquotedbl Relative State\textquotedbl{} Formulation
of Quantum Mechanics. \textit{Rev. Mod. Phys.} 29: 454-462.

18. DeWitt BS, Graham N 1973 The Many-Worlds Interpretation of Quantum
Mechanics. Princeton University Press, Princeton.

19. Zurek WH 2005 Probabilities from entanglement, Born\textquoteright s
rule pk = |\textgreek{y}k|2 from envariance. \textit{Phys. Rev. }A
71, 052105

20. Zurek WH 2007 Relative states and the environment: einselection,
envariance, quantum darwinism, and the existential interpretation.
arXiv:0707.2832 v1

21. Hooft GT 2016 The Cellular Automaton Interpretation of Quantum
Mechanics. Springer: Berlin, Germany.

22. Hooft GT 2014 The cellular automation interpretation of quantum
mechanics. A view on the Quantum Nature of our Universe, compulsory
or i3possible? arXiv:1405.1548.

24. Elze HT 2014 Action principle for cellular automata and the linearity
of quantum mechanics.\textit{ Phys. Rev.} A 89 012111 arXiv:1312.1655.

25. Buniy R, Hsu SDH, Zee A 2006 Discreteness and the origin of probability
in quantum mechanics. arXiv:hep-th/0606062v2. 

26. Einstein A, Podolsky B, Rosen N 1935 Can Quantum-Mechanical Description
of Physical Reality Be Considered Complete? \textit{Phys. Rev. }47:777\textendash 780. 

27. Bell J 1964 On the Einstein Podolsky Rosen Paradox. \textit{Physics}
1:195\textendash 200.

28. Aspect A 1999 Bell's inequality test: more ideal than ever\textit{.
Nature} 398:189\textendash 90.

29. Hense B, Bernien H, Dreau AE, Reiserer A, Kalb N, Blok MS, Ruitenberg
J, Vermeulen RFL, Schouten RN, Abellan C, Amya W, Prueri V, Mitchell
MW, Markham M, Twitchen DJ, Ekouss D, Wehner S, Taminiau TH, Hanson
R 2015 Loophole-free Bell inequality violation using electron spins
separated by 1.3 kilometres. \textit{Nature} 526:682-686.

30. Wittmann B, Ramelow S, Steinlechner F, Langford NK, Brunner N,
Wiseman HM, Ursin R, Zeilinger A 2012 Loophole-free Einstein\textendash Podolsky\textendash Rosen
experiment via quantum steering N. \textit{J. of Physics} 14: 053030-05042.

31. Brune M, Schmidt-Kaler F, Maali A, Dreyer J, Hagley E, Raimond
JM., Haroche S 1996 Quantum Rabi Oscillation: A Direct Test of Field
Quantization in a Cavity \textit{Phys. Rev. Lett}. 76: 1800-1803.

32. Deléglise S, Dotsenko I, Sayrin C, Bernu J, Brune M, Raimond JM,
Haroche S 2008 Reconstruction of non-classical cavity field states
with snapshots of decoherence \textit{Nature} 455:510-514.

33. Penrose R 2004 The road to reality. Random House, London.
\end{document}